\documentclass[prl,aps,twocolumn,superscriptaddress]{revtex4-1}
\usepackage{xcolor}
\usepackage{graphicx,color}
\usepackage{amsthm}
\usepackage{amsfonts}
\usepackage{algorithmic}
\usepackage{enumerate}
\usepackage{latexsym}
\usepackage{amsmath}
\usepackage{amssymb}
\usepackage[colorlinks=true,citecolor=blue,linkcolor=blue]{hyperref}
\def\avg#1{\left\langle#1\right\rangle}

\usepackage{dcolumn}
\usepackage{bm}

\usepackage[T1]{fontenc}
\usepackage{mathptmx}

\begin{document}
\title{Magnetic fluctuation and dominant superconducting pairing symmetry near the tunable Van Hove singularity}

\author{Xiaohan Kong}
\email{These authors contributed equally to this work.}
\affiliation{Department of Physics, Beijing Normal University, Beijing 100875, China\\}
\author{Boyang Wen}
\email{These authors contributed equally to this work.}
\affiliation{Department of Physics, Beijing Normal University, Beijing 100875, China\\}
\author{Kaiyi Guo}
\affiliation{Department of Physics, Beijing Normal University, Beijing 100875, China\\}
\author{Ying Liang}
\affiliation{Department of Physics, Beijing Normal University, Beijing 100875, China\\}
\affiliation{Key Laboratory of Multiscale Spin Physics(Ministry of Education), Beijing Normal University, Beijing 100875, China\\}
\author{Tianxing Ma}
\email{txma@bnu.edu.cn}
\affiliation{Department of Physics, Beijing Normal University, Beijing 100875, China\\}
\affiliation{Key Laboratory of Multiscale Spin Physics(Ministry of Education), Beijing Normal University, Beijing 100875, China\\}

\begin{abstract}
We have investigated the magnetism and pairing correlations of the triangular lattice based on the Hubbard model using the determinant quantum Monte Carlo method and the constrained path Monte Carlo. The results show that the presence of the next-nearest-neighbor hopping integral $t^{\prime}$ introduces an additional energy scale to the system, and through $t^{\prime}$, one can regulate the shape of the density of states and thus the position of the van Hove singularity point. Increasing inverse temperature $\beta$ and on-site interaction $U$ favor the formation of ferromagnetic correlation in a rather large filling region, and the calculations for different lattice sizes show that the range of the ferromagnetic correlations is smaller than the smallest lattice simulated at the investigated temperatures. We study the different pairing correlations of the triangular lattice near several typical fillings and show that the $f$-wave pairing dominates the system in the filling region near the van Hove singularity point with a high density of states, where the ferromagnetic correlation is also enhanced. When the filling is close to half-filling, the pairing susceptibility with $f$ wave is suppressed and the pairing susceptibility of $f_n$ wave is enhanced, however, both the effective pairing interaction with $f$ wave and $f_n$ wave are negative, which indicates that neither $f$-wave nor $f_n$-wave superconductivity may exist. Finally, we find that the pairing channel of different symmetry in the system maybe closely related to the magnetic properties. Ferromagnetic fluctuation favors the formation of $f$-wave pairing, while antiferromagnetic fluctuation tends to promote $f_n$-wave pairing.
\end{abstract}


\maketitle
\section{\romannumeral1. Introduction}

A strongly correlated electron system will possess a partial flat band in the band dispersion due to the peculiarities of the lattice structure, long-range hopping of electrons, etc. In the flat band region, $E(\bf{k})$ hardly varies with $\bf{k}$, which means that a large number of quantum states have similar kinetic energy, and the density of states (DOS) here is higher. At this time, the system is more sensitive to changes in potential energy, which easily induces many exotic quantum phases, such as charge and spin density waves\cite{11111111111,PhysRevB.99.144507} and unconventional superconductivity\cite{Yuan2019,PhysRevB.101.014501}, thus attracting much attention. One famous example is the heavy fermion system of rare-earth metal compounds based on the Kondo lattice$-$ through the exchange between free and bound electrons, a flat band is generated close to the Fermi surface, thus forcing the system to exhibit superconductivity\cite{Ernst2011,Steglich_2016}. The flat band due to strong interlayer coupling near the Fermi energy level in a twisted bilayer graphene superlattice is another typical example, which is generated by the crossing of two energy levels near the Dirac point\cite{Cao2018}, and the appearance of this flat band is considered to be the source of superconductivity. In addition, the kagome lattice, which has recently become a hot topic, is one of the most property-rich lattice structures in the flat band family.
It has been shown that the unconventional superconductivity found in kagome metals may originate from sublattice interference\cite{PhysRevLett.127.177001}, and the presence of its flat band offers the possibility of realizing a ferromagnetic correlation\cite{Mielke_1991,PhysRevLett.100.136404} and a nontrivial topology\cite{PhysRevB.93.155155,PhysRevLett.125.266403}.

Recently, a flat band system based on a triangular lattice was studied by Sayyad $et$ $al$\cite{Sayyad_2023}. They took the next-nearest-neighbor hopping $t^{\prime}$ into account, discussed the electronic nematicity driven by many-body interactions and its interaction with superconductivity, and showed that a stronger nematicity in the flat band contributes to an increase in $T_{c}$ and can drive the pairing form to undergo a transition. Inspired by this, we investigate the magnetism and pairing correlations for the triangular lattice of this system. $t^{\prime}$ not only provides us with a good platform for realizing a flat band but also provides new competing energy scales for geometric frustration with electron correlations. Unlike the square lattice, the triangular lattice has stronger ferromagnetism in the high DOS region near the van Hove singularity (VHS) point due to the asymmetry of the DOS distribution and the higher degeneracy at the single-particle energy level\cite{PhysRevB.35.3359,PhysRevB.62.10033,doi:10.1143/JPSJ.72.2437,PhysRevB.77.125114,PhysRevB.80.014428}. When the on-site interaction $U$ is nonzero, the energy level width at a higher DOS is narrower, thus, the effect of the electron kinetic energy on the total energy is finite, and the effect of $U$ gets larger proportion. When $U$ increases, the energy of the ground state with total spin $S=0$ may be higher than that of the partially polarized state with $S\neq0$. A further increase in $U$ may lead to the appearance of ferromagnetic correlations. For the system we study, we approximate that the interaction $U$ does not have a remarkable effect on the DOS (especially near the Fermi surface). To study superconductivity, we can observe the behavior and trends of pairing correlations.

Scholars have studied the behavior of each pairing channel in the triangular lattice for different scenarios, but no clear conclusions have been given\cite{PhysRevB.63.174507,PhysRevB.68.104510,PhysRevB.77.125114,doi:10.1143/JPSJ.73.17,doi:10.1143/JPSJ.71.2629,PhysRevLett.93.077001,PhysRevB.97.155145,PhysRevLett.128.167002}. In the low electron filling region, perturbation theory calculations based on the Hubbard model show that $d$-wave and $p$-wave pairings are stable\cite{doi:10.1143/JPSJ.73.17,doi:10.1143/JPSJ.71.2629}; however, the results of renormalization group calculations show that ($d+id$)-wave pairing with the antiferromagnetic exchange effect is more stable\cite{PhysRevB.68.104510}. The appearance of unconventional superconductivity in Na$_{x}$CoO$_{2}\cdot y$H$_{2}$O is also explained by the spin-triplet pairing mechanism induced by the disconnected Fermi surface\cite{PhysRevB.63.174507,PhysRevLett.93.077001}. TRILEX method results show that the $d$-wave superconducting phase appears in the triangular lattice adatom system based on the Si(111) surface with long-range interactions when doped with holes\cite{PhysRevB.97.155145}. Meanwhile, in a hole-doped triangular lattice consisting of Sn atoms deposited on a Si substrate, weakly coupled renormalization group calculations show that $d$-wave pairing is suppressed in the vicinity of the VHS point when nonlocal interactions are considered, and instead, chiral $p$-wave competition with the $f$-wave is exhibited\cite{PhysRevLett.128.167002}.

In this work, we studied the various behaviors of magnetic and pairing correlations for the system mainly at $t^{\prime} = -0.15$, and our focus of interest is mainly around the VHS point and in proximity to half-filling. We investigated the magnetic correlation of the system by calculating the spin susceptibility and studied the pairing unequal time susceptibility of the system under a finite temperature and the long-range pairing correlation function at zero temperature using the determinant quantum Monte Carlo (DQMC) method. We calculated the energies of different spin subspaces to infer the appearance of the ferromagnetic ground state using the constrained path Monte Carlo (CPMC) method. 

\begin{figure}[t]
\includegraphics[scale=0.5]{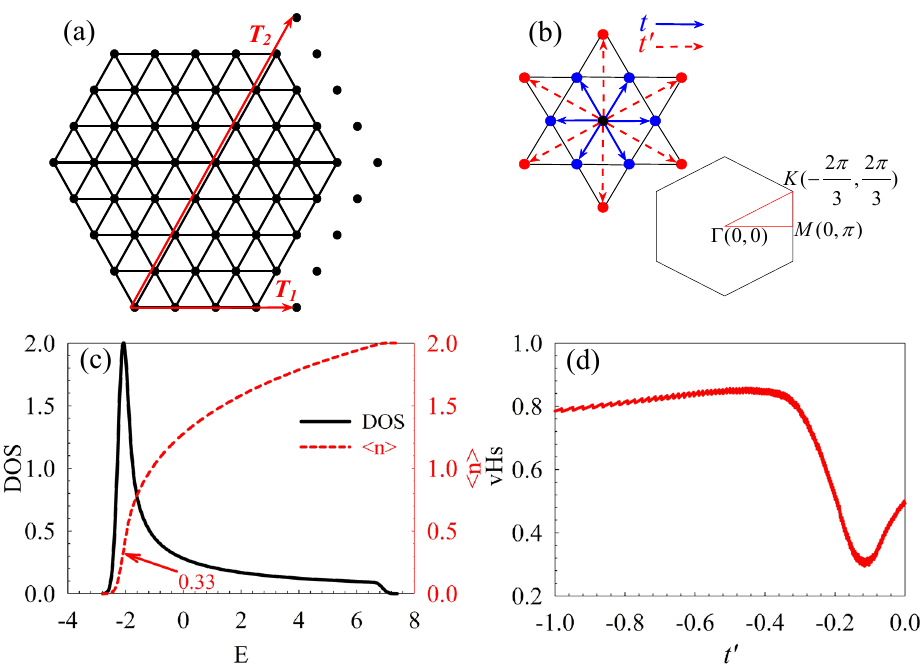}
\caption{\label{Fig1} (a) Schematic diagram of the triangular lattice with $L = 4$. $\boldsymbol{T_1}$ and $\boldsymbol{T_2}$ represent the translation vectors of the real-space lattice. There are $3\times L^2$ sites in total and $2L$ sites along the diagonal. (b) The hopping amplitudes are $t$ for the six nearest neighbors (blue solid arrow) and $t^{\prime}$ for the six next nearest neighbors (red dashed arrow). The hexagonal first Brillouin zone (BZ) of the triangular lattice is marked with representative points along the high symmetry path, where $\Gamma(k_x,k_y)=(0,0)$, $M(k_x,k_y)=(0,\pi)$, and $K(k_x,k_y)=(-2\pi/3,2\pi/3)$. (c) DOS (dark solid line) and filling $\langle n \rangle$ (red dashed line) as a function of energy with $t^{\prime} = -0.15|t|$. (d) Dependence of the position of the VHS point on the next-nearest-neighbor hopping amplitude $t^{\prime}$.}
\end{figure}

\begin{figure}[t]
\includegraphics[scale=0.3]{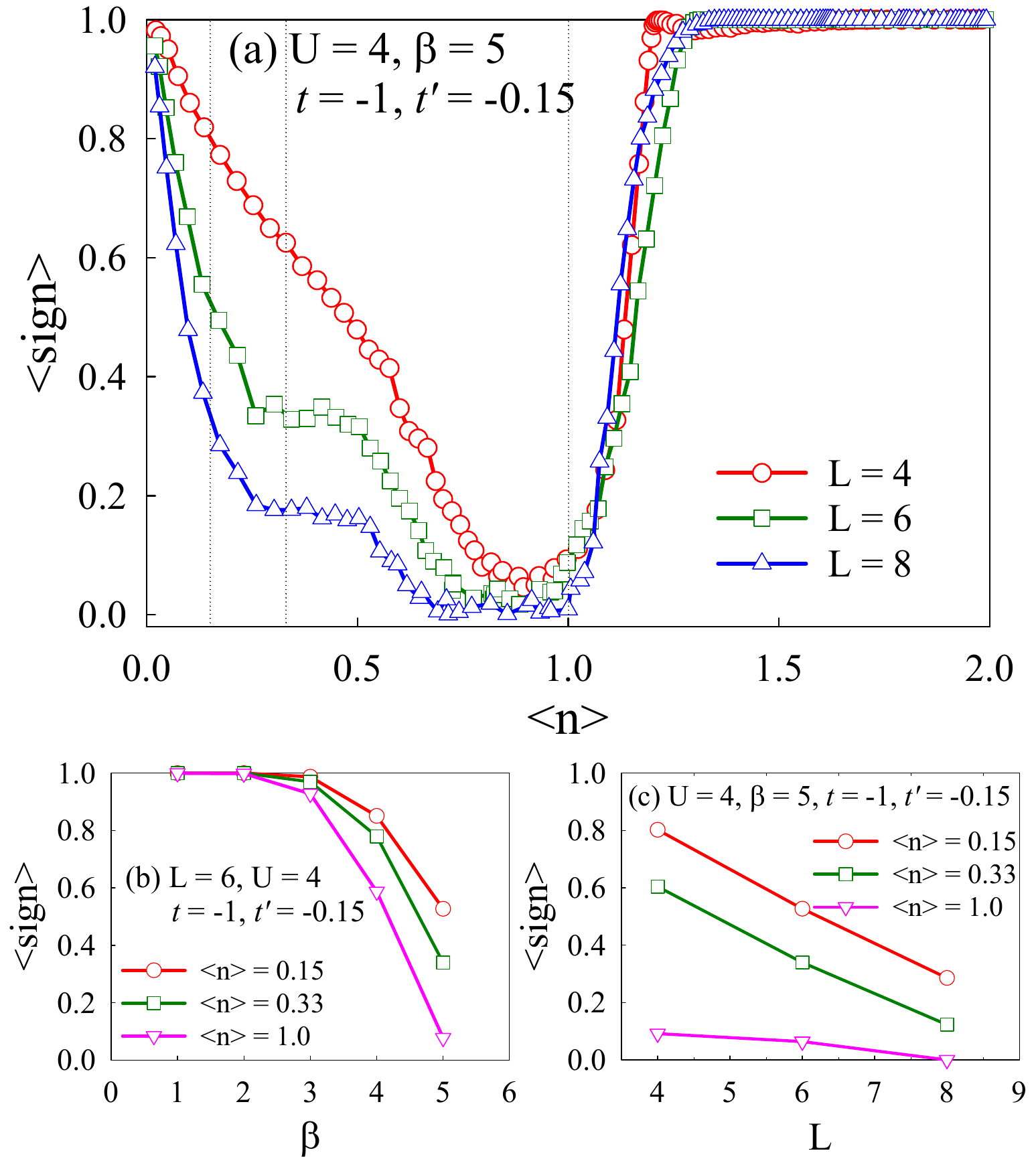}
\caption{\label{Fig2_SP} "Sign problem" $\langle \textup{sign} \rangle$ for the triangular lattice as a function of (a) filling $\langle n \rangle$ with different $L$ values at $\beta = 5$, (b) $\beta$ with several fixed fillings for $L = 6$ and (c) $L$ with several fixed fillings at $\beta = 5$. All of the cases are fixed at $U=4$, $t=-1$, and $t^{\prime}=-0.15$. The fillings used in panels (b) and (c) are marked by dark dotted lines in (a).}
\end{figure}
\begin{figure*}[t]
\includegraphics[width=1.0\textwidth,height=0.28\textwidth]{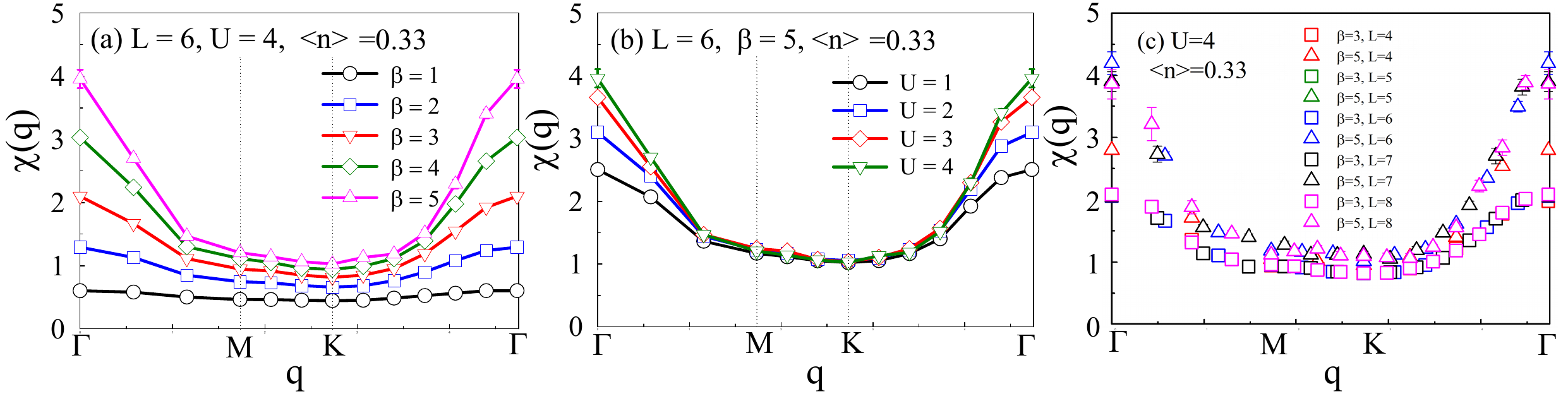}
\caption{\label{Fig3_bUs} Magnetic susceptibility $\chi(\pmb{q})$ versus $\pmb{q}$ for (a) various values of the inverse temperature $\beta$ with $L=6$ and $U=4$, (b) various values of the interaction $U$ at $L=6$ and $\beta = 5$ and (c) various values of $L$ at $\beta =3, 5$ and $U = 4$. The filling is fixed at $\langle n \rangle = 0.33$ for all the cases, which corresponds to the VHS point for $t^{\prime} = -0.15$. Data are shown along the high symmetry path $\Gamma \rightarrow M \rightarrow K \rightarrow \Gamma$ in the hexagonal first BZ, and the dark dotted lines mark the positions of $M$ and $K$.}
\end{figure*}

\section{\romannumeral2. MODEL}
The triangular lattice can be well described by the single-band Hubbard model,
\begin{eqnarray}
\label{Hamiltonian}
\hat{H}&=&-t\sum_{\langle\bf{i,j}\rangle\sigma}(\hat{c}_{\bf{i}\sigma}^\dagger\hat{c}_{\bf{j}\sigma}
+H.c.)-t^{\prime}\sum_{\left[\bf{i,j}\right]\sigma}(\hat{c}_{\bf{i}\sigma}^\dagger\hat{c}_{\bf{j}\sigma}
+\textup{H.c.}) \nonumber\\
&&+U\sum_{\bf{i}}\hat{n}_{\bf{i}\uparrow}\hat{n}_{\bf{i}\downarrow}-\mu\sum_{\bf{i}\sigma}\hat{n}_{\bf{i}\sigma},
\end{eqnarray}
where $t$ and $t^{\prime}$ are the nearest-neighbor (NN) and next-nearest-neighbor (NNN) hopping integrals, respectively, $U$ is the on-site Coulomb interaction, and $\mu$ is the chemical potential. Here, $\hat{c}_{{\bf i}\sigma}^\dagger$ $(\hat{c}_{{\bf i}\sigma})$ annihilates (creates) a particle at site ${\bf i}$ with spin $\sigma$ $(\sigma = \uparrow,\downarrow)$, and $\hat{n} = c_{{\bf i}\sigma}^\dagger c_{{\bf i}\sigma}$. The signs of $t$ and $t^{\prime}$ can be positive or negative, corresponding to the electron and hole pictures, respectively. The sign has no specific physical meaning since both signs can be related by particle-hole transformations: $c_{\bf{i}}\dagger\leftarrow d_{\bf{i}}$. In the rest of this paper, we take $\left|t\right| = 1$ as the energy unit and $t, t^{\prime}<0$, which means that the Hamiltonian describes the hole behavior of the system. Our main numerical simulations were performed on the sketch of the triangular lattice with a hexagonal shape shown in Fig.~\ref{Fig1}(a). There are $2L$ sites along the diagonal, and the number of sites in this series of lattices is 3$\times L^2$. This lattice structure is set up such that most of the geometric symmetry of the triangular lattice is preserved. The data points of the first Brillouin zone (BZ) include all the high symmetry points, such as $\Gamma$, $M$, and $K$, as shown in Fig.~\ref{Fig1}(b). We give the noninteracting ($U = 0$) band dispersion $E(k_x,k_y)$ for the triangular lattice as
\begin{eqnarray}
E(k_x,k_y)=-t[2\cos(k_x)+4\cos(k_x/2)\cos(\sqrt{3}k_y/2)] \nonumber \\
-t^{\prime}[2\cos(\sqrt{3}k_y)+4\cos(3k_x/2)\cos(\sqrt{3}k_y/2)].
\end{eqnarray}

In contrast to the ideal triangular Hubbard model with only nearest-neighbor hopping, the presence of long-range hopping $t^{\prime}$ delocalizes the particle and further competes with the geometrical frustration, which causes the VHS point to move, as shown in Figs.~\ref{Fig1}(c) and (d). The high DOS at the VHS point contributes to enhancement of the ferromagnetic spin fluctuation in this particle-filling region. To study the magnetism on this lattice, we define the spin susceptibility in the $z$ direction at zero frequency ($\omega = 0$) as
\begin{eqnarray}
\chi(\pmb{q})=\int_{0}^{\beta}d\tau\sum_{\bf{i,j}} e^{i{\bf q}\cdot (\bf i-\bf j)}\langle m_{\bf i}(\tau)\cdot m_{\bf j}(0)\rangle,
\end{eqnarray}
where $m_{\pmb{i}}(\tau) = e^{H\tau}m_{\pmb{i}}(0) e^{-H\tau}$, with $m_{\pmb{i}} \equiv c_{\pmb{i}\uparrow}^\dagger c_{\pmb{i}\uparrow}-c_{\pmb{i}\downarrow}^\dagger c_{\pmb{i}\downarrow}$. Here, $\chi$ is measured in units of $\left|t\right|^{-1}$, and $\chi(\Gamma)$ quantifies the ferromagnetic correlation, while $\chi(K)$ quantifies the antiferromagnetic correlation.

\begin{figure*}[t]
\includegraphics[width=1.0\textwidth,height=0.29\textwidth]{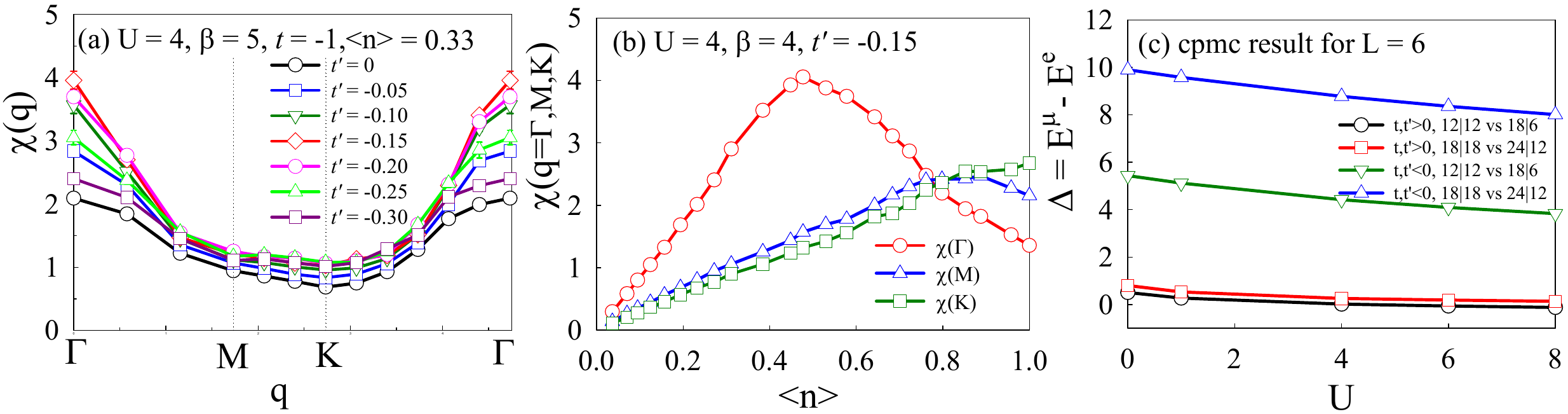}
\caption{\label{Fig4} (a) Magnetic susceptibility $\chi(\pmb{q})$ versus $\pmb{q}$ for various values of $t^{\prime}$ at $L=6$, $U = 4$, $\beta = 5$, $t = -1$, and $\avg{n}= 0.33$. (b) Magnetic susceptibility at $\Gamma$, $M$, and $K$ as a function of filling $\avg{n}$ at $U = 4$, $\beta = 5$ and $t^{\prime} = -0.15$. (c) Ground state energy difference $\Delta=E_{0}^{\mu}-E_{0}^{e}$ between $S\neq 0$ and $S = 0$ for different closed shell cases.}
\end{figure*}

\section{\romannumeral3. RESULTS AND DISCUSSION}
Before investigating the magnetism and superconductivity of this system, we first studied the sign problem of the DQMC method to select the appropriate $U$, $\beta$, and filling regions for numerical simulations. It is easy to see from Fig.~\ref{Fig2_SP} that the more severe frustration brought about by the existence of $t^{\prime}$ and its competition with electron correlations make the sign problem of the system more serious than when only nearest-neighbor hopping is considered\cite{PhysRevB.92.045110}.
In our simulations, 3000 $\sim$ 12 000 sweeps were used to equilibrate the system, and an additional
10 000 $\sim$ 160 000 sweeps were then performed, each of which
generated a measurement. These measurements were divided
into 10 $\sim$ 20 bins that provide the basis of coarse-grain averages,
and errors were estimated based on standard deviations from
the average.  Indeed, to obtain the same quality of data as
$\avg{\textup{sign}}\sim 1$, much longer runs are necessary to compensate
the fluctuations as $\avg{\textup{sign}}< 1$, and we can estimate that the runs need
to be stretched by a factor on the order of $\avg{\textup{sign}}^{-2}$\cite{santos2003introduction,PhysRevB.104.035104,PhysRevB.105.045132,PhysRevB.101.155413,PhysRevB.106.205149}.
Based on this, subsequent studies were carried out with the appropriate parameters.

The behavior of the spin susceptibility does not fundamentally change in the studied filling regions, so most of our results are focused on hole filling $\langle n \rangle = 0.33$, which corresponds to the VHS point in this system. Figure~\ref{Fig3_bUs}(a) shows $\chi(\pmb{q})$ as a function of $\pmb{q}$ at different values of the inverse temperature $\beta$ for $U = 4$ on a lattice of $L = 6$. We can see that $\chi(\pmb{q})$ exhibits a wide peak around the $\Gamma$ point in the BZ, while the minimum value is located near the $K$ point, indicating some ferromagnetic fluctuation in the system. We can also see that the peak around the $\Gamma$ point increases more slowly with decreasing temperature in the inverse temperature range of $3\leq\beta\leq5$. However, due to the limitation of the sign problem of the DQMC method, we are unable to perform numerical calculations for the cases at lower temperatures ($\beta>5$), and there are no definite results about the low-temperature behavior of $\chi(\pmb{q})$. Figure~\ref{Fig3_bUs}(b) shows the results of the relationship between $\chi(\pmb{q})$ and $\pmb{q}$ for different on-site interactions $U$ at $\beta=5$. At the same temperature, the peak at the $\Gamma$ point increases with larger $U$, indicating that the on-site interaction $U$ tends to strengthen the ferromagnetic fluctuations. It is worth noting that this strengthening effect of $U$ is not present at the K point of the BZ, and $\chi$(K) quantifies the antiferromagnetic correlation.

In Fig.~\ref{Fig3_bUs}(c), we show the effect of different lattice sizes, corresponding to $L=4$, $5$, $6$, $7$, and $L=8$, on the behavior of $\chi(\pmb{q})$ for $\beta=3$ and $\beta=5$ with $U=4|t|$. Except $L=4$ at $\beta=5$, the data obtained at different lattice sizes are consistent within the statistical error allowed, i.e., the $\chi(\pmb{q})$ obtained by summing over all the sites does not vary with the lattice size. Due to the limitation of QMC method, lower temperature and larger lattice size are inaccessible, and we argue here that the range of the ferromagnetic correlations is smaller than the smallest lattice shown in Fig.\ref{Fig3_bUs}(c).

Figure~\ref{Fig4}(a) shows the effect of different next-nearest-neighbor hopping $t^{\prime}$ values on $\chi(\pmb{q})$ for $U = 4|t|$ and $\beta = 5$, with the filling fixed at $\avg{n} = 0.33$. Focusing our attention on the $\Gamma$ point in the BZ, we can see that $\chi(\Gamma)$ first increases as $|t^{\prime}|$ increases from 0 and then exhibits a decreasing behavior across $|t^{\prime}| = 0.15$, i.e., the maximum value of $\chi(\Gamma)$ appears at $t^{\prime} = -0.15$. It is worth noting that the filling $\avg{n}=0.33$ we study is exactly the VHS point of the system at $t^{\prime} = -0.15$. The results show that the high DOS favors the ferromagnetic fluctuation of the system. This can also be demonstrated in another way, as shown in Fig.~\ref{Fig4} (b), which shows the behavior for different filling regions $\chi(\pmb{q})$ at $U = 4|t|$ with $\beta$ fixed at 4. The ferromagnetic fluctuations are enhanced in the noninteracting case when the filling is close to the region with a higher DOS, especially around $\avg{n}=0.33$, where $\chi(\Gamma)$ grows most rapidly. This indicates that a higher DOS in this system is favorable for promoting the formation of ferromagnetic correlations. Furthermore, the $\chi(\pmb{q})$ fluctuation becomes stronger as $\avg{n}$ increases, and the peak tends to move toward the $M$ point due to the competition between ferromagnetic and antiferromagnetic fluctuations. At approximately $\avg{n} \approx 0.78$, the competition between ferromagnetic and antiferromagnetic fluctuations reaches its peak. Previous studies have shown that the antiferromagnetic correlation is strong near the half-filling region, where $\chi(\pmb{q})$ peaks at the $K$ point, and dominates over a considerable range until close to the VHS point\cite{https://doi.org/10.1002/andp.19955070405,PhysRevLett.95.037001}. In this range, the presence of ferromagnetism will suppress the antiferromagnetism to some extent, causing the peak of $\chi(\pmb{q})$ to be shifted toward the $M$ point located between $\Gamma$ and $K$. This is consistent with the conclusions given by mean-field theory\cite{https://doi.org/10.1002/andp.19955070405} and perturbation theory\cite{doi:10.1143/JPSJ.73.17}. Thus, our DQMC-based results suggest that a high DOS is crucial for the ferromagnetic fluctuations of this system.

\begin{table}[t]
\centering
\caption{\label{table1} Comparison of ground state energies in spin subspaces for different combinations of particles. $U = 0$ corresponds to the noninteraction case, $t, t^{\prime}>0$ indicates that the particles are electrons, and $t, t^{\prime}<0$ indicates holes. $\Delta=E_{0}^{u}-E_{0}^{e}$ represents the energy difference between two spin subspaces. $E_{0}^{u}$ denotes a system with unequal numbers of spin-up and spin-down particles, while $E_{0}^{e}$ denotes a system with equal numbers of both.}
	\begin{ruledtabular}
		\begin{tabular}{cccccc}
			$U$&$0$&$1|t|$&$4|t|$&$6|t|$&$8|t|$\\
			\hline\\
			\multicolumn{6}{c}{$t,t^{\prime}>0$}\\\\
			24 particles (18$\uparrow$ 6$\downarrow$) & \textendash52.72 & \textendash52.02 & \textendash51.27 & \textendash51.02 & \textendash50.83 \\
			24 particles (12$\uparrow$ 12$\downarrow$) & \textendash53.22 & \textendash52.29 & \textendash51.29 & \textendash50.96 & \textendash50.71 \\
        $\Delta=E_{0}^{\mu}-E_{0}^{e}$ & 0.50 & 0.27 & 0.02 & \textendash0.06 & \textendash0.10 \\\\

			36 particles (24$\uparrow$ 12$\downarrow$) & \textendash77.47 & \textendash75.52 & \textendash73.31 & \textendash72.58 & \textendash72.04 \\	
			36 particles (18$\uparrow$ 18$\downarrow$) & \textendash78.27 & \textendash76.05 & \textendash73.57 & \textendash72.77 & \textendash72.18 \\
        $\Delta=E_{0}^{\mu}-E_{0}^{e}$ & 0.80 & 0.53 & 0.26 & 0.19 & 0.14 \\\\
			\multicolumn{6}{c}{$t,t^{\prime}<0$}\\\\

        24 particles (18$\uparrow$ 6$\downarrow$) & \textendash120.56 & \textendash119.63 & \textendash117.47 & \textendash116.41 & \textendash115.54 \\
			24 particles (12$\uparrow$ 12$\downarrow$) & \textendash125.99 & \textendash124.74 & \textendash121.89 & \textendash120.50 & \textendash119.37 \\
        $\Delta=E_{0}^{\mu}-E_{0}^{e}$ & 5.43 & 5.11 & 4.42 & 4.09 & 3.83 \\\\

    		36 particles (24$\uparrow$ 12$\downarrow$) & \textendash158.51 & \textendash156.02 & \textendash150.30 & \textendash147.52 & \textendash145.29 \\
			36 particles (18$\uparrow$ 18$\downarrow$) & \textendash168.41 & \textendash165.60 & \textendash159.08 & \textendash155.88 & \textendash153.30 \\
        $\Delta=E_{0}^{\mu}-E_{0}^{e}$ & 9.90 & 9.58 & 8.78 & 8.36 & 8.01 \\
	\end{tabular}
	\end{ruledtabular}
\end{table}

On the other hand, we can also use the CPMC method to predict the emergence of the ferromagnetic ground state by calculating the energy difference of different spin subspaces, and the numerical results are listed in Table ~\ref {table1}. To unify the results with those of the DQMC method and simplify the simulations, here, we describe the system in terms of the electron picture for high hole filling with $t, t^{\prime}>0$ and electron filling $2-\avg{n}$. We studied the following four cases on a triangular lattice with $L=6$: (i) 18 electrons with spin up and six electrons with spin down versus 12 up and 12 down for a total of 24 electrons, corresponding to hole filling $\avg{n}\approx 1.78$; (ii) 24 electrons with spin up and 12 electrons with spin down versus 18 up and 18 down for a total of 36 electrons, corresponding to hole filling $\avg{n} \approx 1.67$; (iii) 18 holes with spin up and six holes with spin down versus 12 up and 12 down for a total of 24 holes, corresponding to hole filling $\avg{n} \approx 0.22$; and (iv) 24 holes with spin up and 12 holes with spin down versus 18 up and 18 down for a total of 36 holes, corresponding to hole filling $\avg{n} \approx 0.33$. From Table ~\ref {table1} and Fig.~\ref{Fig4}(c), it can be seen that the energy difference $\Delta=E_{0}^{u}-E_{0}^{e}$ between the paramagnetic ground state and the partially polarized state decreases with increasing on-site interaction $U$, and the ferromagnetic instability becomes more pronounced, indicating the contribution of $U$ to promoting the formation of ferromagnetic fluctuations. From this figure, we can clearly see that compared to the lower filling region, the energy difference in the higher filling region is smaller. Meanwhile, compared to other filling regions, the ferromagnetic fluctuations at the VHS point are more significant, and the energy levels are also closer. These results are consistent with our previous research.

\begin{figure}[t]
\includegraphics[width=0.45\textwidth,height=0.6\textwidth]{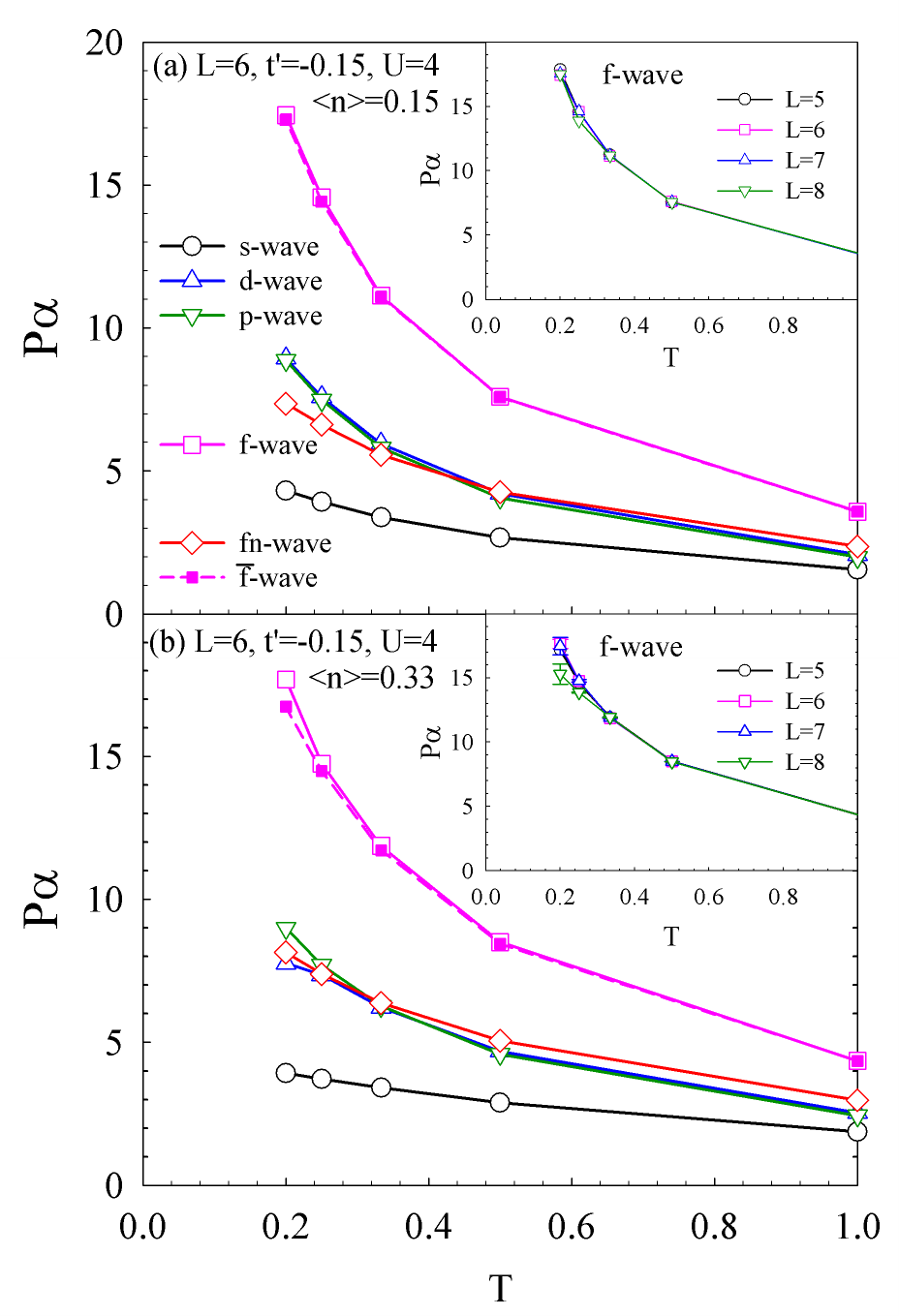}
\caption{\label{Fig6} Pairing susceptibility $P_{\alpha}$ of different pairing symmetries and single particle contribution of $f$-wave as a function of temperature $T$. The insets show the pairing susceptibility of different sites. Calculations are performed at $U = 4$ with (a): $\avg{n} = 0.33$ and (b): $\avg{n}= 0.15$. }
\end{figure}

In line with the previous investigation, we study the pairing correlation for different channels with several typical fillings. First, we can define the order parameter operator as\cite{PhysRevB.37.5070,PhysRevB.37.7359}:
\begin{equation}\label{eq4}
\begin{aligned}
\Delta_{\alpha}^{\dag}(i)=\sum_{l}f_{\alpha}(l)(c_{i\uparrow}c_{i+l\downarrow}- c_{i\downarrow}c_{i+l\uparrow})^{\dag},
\end{aligned}
\end{equation}
where the summation is over all sites $l$ and neighboring sites $i$, with $f_{\alpha}(l)$ being the form factor of the pairing determined by the lattice (see Appendix). In a system of finite size, we are concerned about the calculation of correlation functions. Observing the behavior and trend of the time-dependent pairing susceptibility at finite temperatures is an effective way to study pairing interactions\cite{PhysRevB.37.5070,PhysRevB.37.7359,PhysRevB.39.7259,PhysRevB.52.16155}, where the pairing susceptibility at zero frequency ($\omega=0$) is defined as
\begin{equation}\label{eq5}
\begin{aligned}
P_{\alpha}=\frac{1}{N_{s}}\sum_{i,j}\int_{0}^{\beta}d\tau\langle\Delta_{\alpha}^{\dag}(i,\tau)\Delta(j,0)\rangle.
\end{aligned}
\end{equation}

In the regions we mainly studied (here, we discuss two particular fillings, $\avg{n} = 0.33$ and $\avg{n}  = 0.15$), the behaviors of the various pairing susceptibilities essentially do not vary. Figure~\ref {Fig6}(a) corresponds to the case of $\avg{n}  = 0.33$; it can be seen that the pairing susceptibility tends to increase with decreasing temperature for all channels, and $P_{f}$ increases the fastest. Similar to the former figure, Fig.~\ref {Fig6}(b) shows the results for the $\avg{n}  = 0.15$ case. It can be seen that $P_{p}$ and $P_{f}$ of the spin triplet continue to increase; in particular, $P_{f}$ still increases the fastest, while $P_{s}$, $P_{d}$, and $P_{f_n}$ seem to be close to saturation. However, due to the sign problem of the DQMC method, our numerical simulations can only be performed up to $T = 0.2|t|$, so it is unclear whether the susceptibility of each pairing continues to increase at lower temperatures. The $P_{\alpha}$ here includes the single-particle effect $\langle c_{i\uparrow}^{\dag}c_{j\uparrow} \rangle\langle c_{i+l\downarrow}^{\dag}c_{j+l^{\prime}\downarrow} \rangle$, which does not contribute to particle-particle interactions. Therefore, we evaluated intrinsic pairing  susceptibilities in different channels by extracting single particle contributions $\overline{P}_{\alpha}(i,j)$\cite{PhysRevB.72.134513,PhysRevB.40.506} and replaced $\langle c_{i\uparrow}^{\dag}c_{j\uparrow}c_{i+l\downarrow}^{\dag}c_{j+l^{\prime}\downarrow}\rangle$ with $\langle c_{i\uparrow}^{\dag}c_{j\uparrow} \rangle\langle c_{i+l\downarrow}^{\dag}c_{j+l^{\prime}\downarrow} \rangle$ in Eq.~\ref {eq4}. It is defined as $\bm{P}_{\alpha}=P_{\alpha}-\overline{P}_{\alpha}$. From Figs.~\ref {Fig6}(a) and \ref {Fig6}(b), we can clearly see that the temperature-dependent behavior of $\overline{P}_{\alpha}$ is very similar to that of $P_{\alpha}$, and the difference between the two indicates an effective pairing effect. At the same time, we also see that the effective pairing effect at $\avg{n}=0.33$ is stronger than that at $\avg{n}=0.15$, which indicates that the enhancement of ferromagnetic fluctuations near the VHS point is conducive to the formation of $f$-wave pairing, so we suspect that there may be a correlation between the magnetism and pairing. The inset of Fig.~\ref {Fig6} shows the effect of different lattice sizes ($L = 5, 6, 7$, and $8$) with $\avg{n}=0.33$ and $\avg{n}=0.15$ on the behavior of $P_{f}$. The data obtained at different lattice sizes are consistent within the statistical error allowed.

\begin{figure}[t]
\includegraphics[width=0.45\textwidth,height=0.35\textwidth]{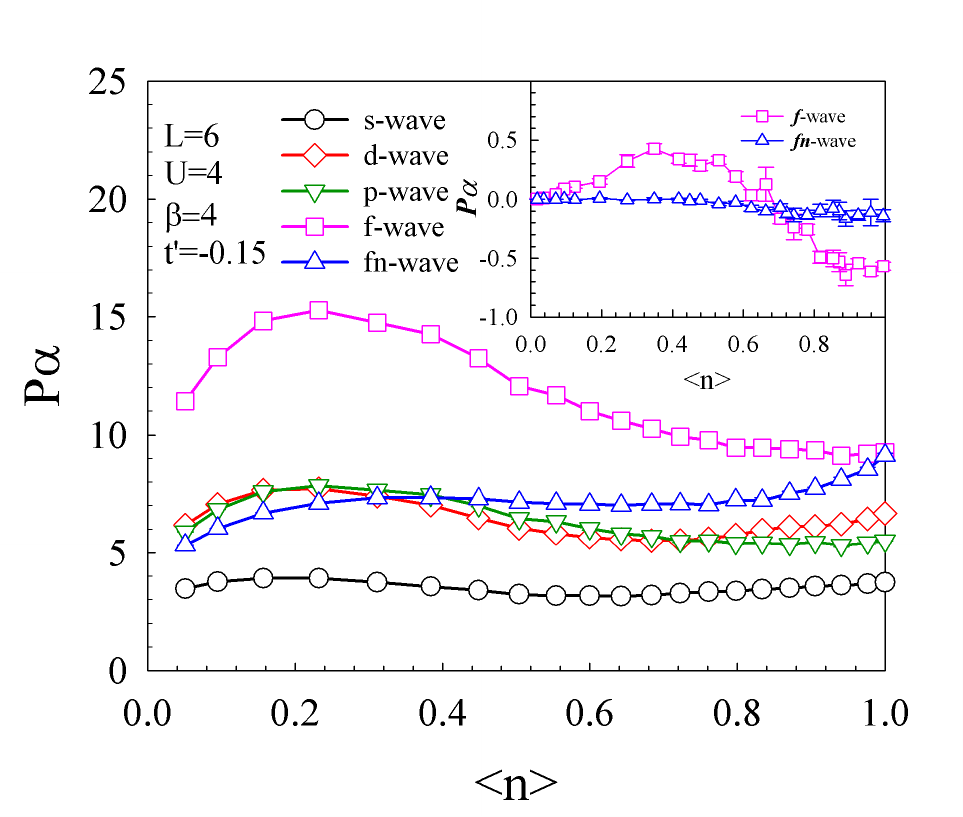}
\caption{\label{Fig7} Pairing susceptibility $P_{\alpha}$ and intrinsic pairing susceptibility $\bm{P}_{\alpha}$(in the inset) of different pairing symmetries as a function of filling $\avg{n}$ at $L=6$, $U=4$, $\beta=4$, and $t'=-0.15$.}
\end{figure}

\begin{figure}[t]
\includegraphics[width=0.45\textwidth,height=0.6\textwidth]{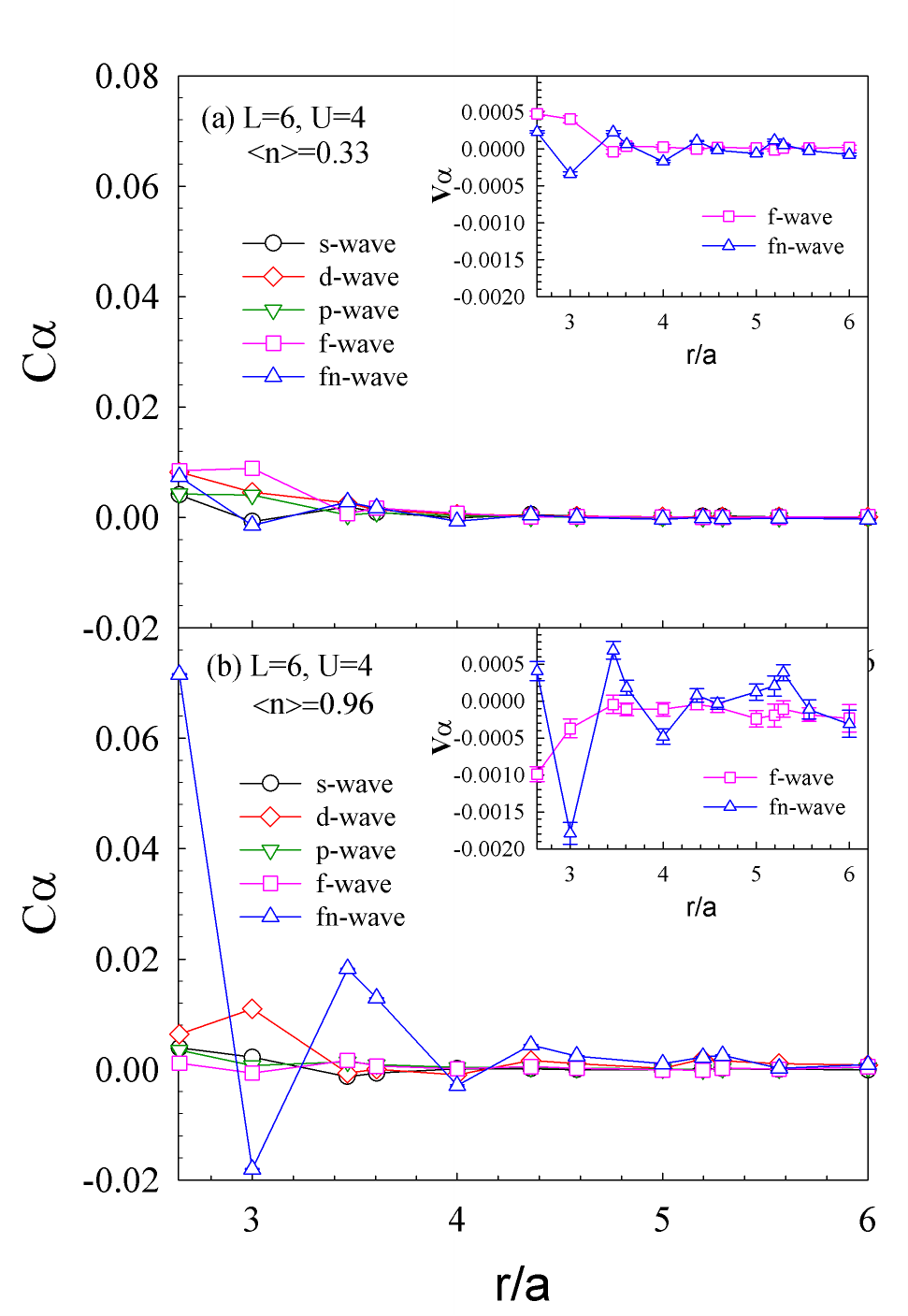}
\caption{\label{Fig8} Long-range correlation of different pairing symmetries as a function of $r/a$. The insets show the vertex contribution as a function of $r/a$. Calculations are performed at $L = 6$, $U = 4$, and $t^{\prime}=-0.15$ with (a) $\avg{n}=0.33$ and (b) $\avg{n}=0.89$.}
\end{figure}

To study the effect of doping on pairing, we calculated the $P_{\alpha}$ for the triangular lattice of $L=6$ and give the phase diagram of the dominant pairing channel at different fillings in Fig.~\ref {Fig7}, where $U=4$ and $\beta = 4$. The $f$-wave pairing dominates in a large filling region, especially near the VHS point.
According to the results of previous studies and Fig.~\ref{Fig4}(b), we know that the system shows ferromagnetic correlation. This is consistent with the previous findings on Na$_{x}$CoO$_{2}\cdot y$H$_{2}$O based on symmetry considerations of the existence of a spin-triplet superconducting mechanism around the filling near the VHS point\cite{PhysRevLett.91.257006}. When $\avg{n}$ is close to half-filling, where the system shows antiferromagnetic correlation, the behavior of pairing susceptibility with $f_n$ wave shows a rapid enhancement and tends to dominate.
In the inset, we plot the intrinsic pairing interaction ${\bm{P}}_{f}$ and ${\bm{P}}_{f_n}$. When $\avg{n}<0.6$, it is clear to see that ${\bm{P}}_{f}$ is positive and increases when close to VHS point. The positive intrinsic pairing interaction indicates that there actually exists attraction for $f$-wave pairing. When close to half filling, both ${\bm{P}}_{f}$ and ${\bm{P}}_{f_n}$ are negative, which indicates that both $f$ and $f_n$ superconducting state may not exist\cite{PhysRevB.72.134513,negative}.

\begin{figure}[t]
\includegraphics[width=0.45\textwidth,height=0.6\textwidth]{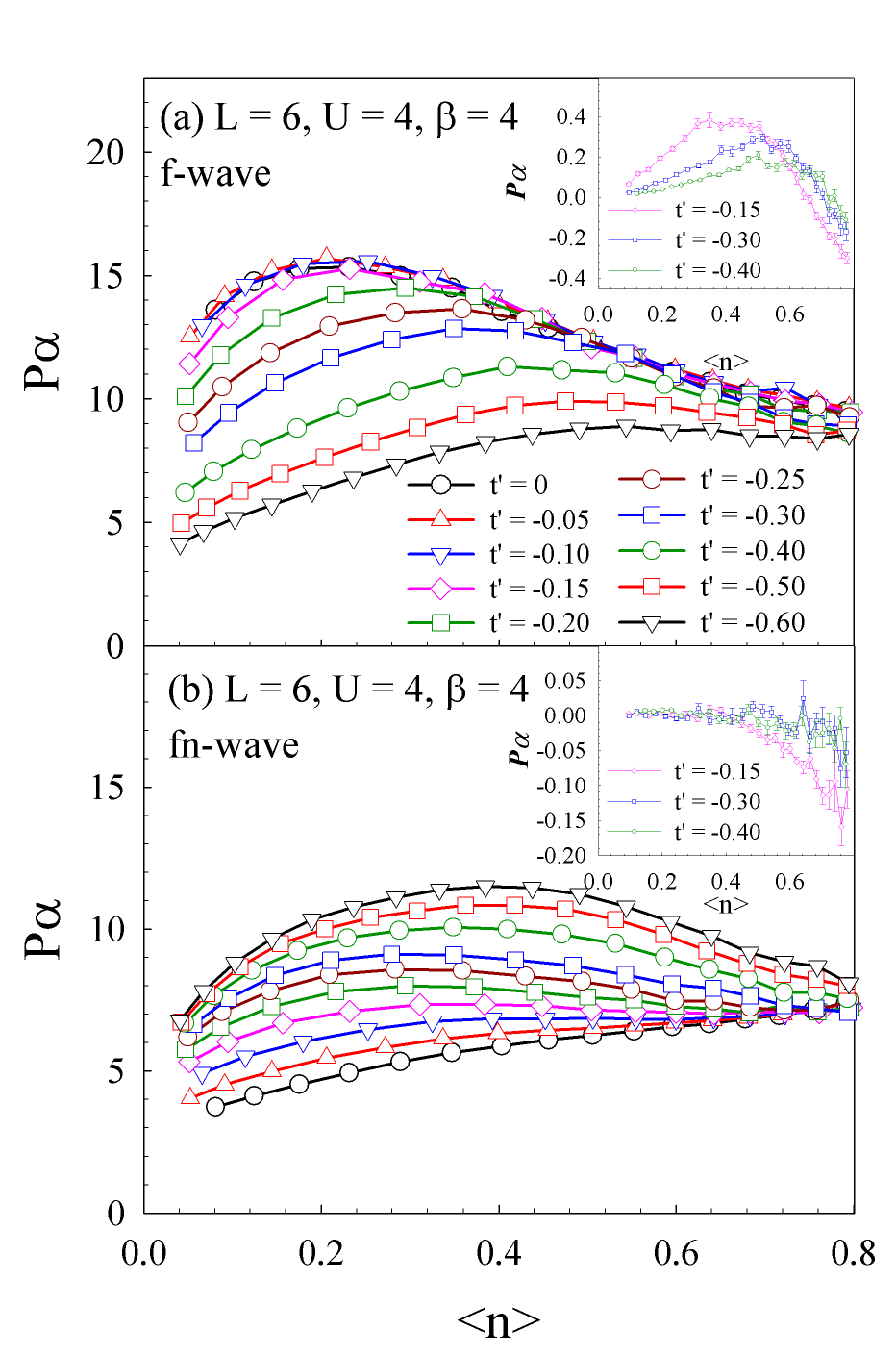}
\caption{\label{Fig9} Pairing susceptibility as a function of filling $\avg{n}$ for different next-nearest-neighbor hopping $t^{\prime}$ values. Calculations are performed at $L=6$, $U = 4$, and $\beta=4$. The insets show intrinsic pairing susceptibility ($\bm{P}_{\alpha}$) as a function of filling $\avg{n}$ for different $t'$.  (a) and (b) present the behavior of $P_{f}$ and $P_{f_n}$ respectively.}
\end{figure}

For the further case at zero temperature, we observed the long-range pairing correlation function, which is defined as\cite{Cheng2014}
\begin{equation}
\begin{aligned}
C_{\alpha}(\bm{r}=\bm{R_{\bf{i}}}-\bm{R_{\bf{j}}})=\langle\Delta_{\alpha}^{\dag}(i)\Delta_{\alpha}(j)\rangle.
\end{aligned}
\end{equation}
We also observed the vertex contributions to the correlations defined by:
	\begin{equation}
		V_{\alpha}({\bm{r}})=C_{\alpha}({\bm{r}})-\overline{C_{\alpha}}({\bm{r}})
	\end{equation}
	where $\overline{C_{\alpha}}({\bm{r}})$ indicates the uncorrelated pairing correlation. For each term in $C_{\alpha}({\bm{r}})$ like $\langle c_{\uparrow}^{\dag}c_{\uparrow}c_{\downarrow}^{\dag}c_{\downarrow}\rangle$, it has a term like $\langle c_{\uparrow}^{\dag}c_{\uparrow}\rangle\langle c_{\downarrow}^{\dag}c_{\downarrow}\rangle$.
The typical fillings of our study are fixed at $\avg{n}=0.33$ and $\avg{n}=0.96$. As seen in Fig.~\ref {Fig8}, when the filling is located near the VHS point ($\avg{n}=0.33$), $C_{f}(\bm{r})$ is overall stronger than other forms of pairing symmetries in the entire long-range interval; with the increase in the filling to $\avg{n}= 0.96$ (near half-filling), $C_{f_n}(r)$ has a greater advantage and tends to dominate. In the inset of Fig.~\ref {Fig8}(a), the vertex contribution of $f$ wave and $f_n$ wave shows similar behavior as pairing correlation. The $f$-wave pairing dominates over the $f_n$-wave pairing near the VHS point at $\avg{n}=0.33$. When electron filling increases to nearly half filling, $V_{f_n}({\bm{r}})$ become larger and $V_{f}({\bm{r}})$  is suppressed.

Based on the above, it is clear that the magnetism is closely related to the pairing correlation. At various fillings, the competition between ferromagnetic and antiferromagnetic correlation plays an important role in the selection of the dominant pairing channel. $(0,0)$ ferromagnetic fluctuation supports $f$-wave pairing, while $(-2\pi/3,2\pi/3)$ antiferromagnetic fluctuation supports $f_n$-wave pairing. For example, in the iron-based superconducting $S_{4}$ model, the $(0,\pi)$ ferromagnetic fluctuation favors the $s_{xy}$-wave electron pairing, while the $(\pi,\pi)$ ferromagnetic fluctuation favors the $d_{x^{2}-y^{2}}$-wave pairing\cite{PhysRevLett.110.107002}. Therefore, our study suggests that there is a close correlation between the magnetism and pairing. At the same time, the results of neutron scattering experiments on high-temperature superconducting iron-phosphorus compounds\cite{Wang2013} and high-pressure nuclear magnetic resonance studies on over-doped iron-phosphorus superconductors\cite{PhysRevLett.111.107004} are consistent with our study, which verifies our conclusion experimentally.

From the previous discussion, we know that the next-nearest-neighbor hopping $t^{\prime}$ can be used to adjust the position of the VHS point, which has a strong "strengthening" effect on the ferromagnetic fluctuation. Because of this, we may indirectly control the magnetism through $t^{\prime}$ to realize selection of the dominant pairing channel of the system. As seen from Fig.~\ref {Fig9}(a), $P_{f}$ shows a weakening trend with increasing $|t^{\prime}|$, and the extreme point slightly shifts to the left before $|t^{\prime}|=0.2$; then, it shifts to the right with larger $|t^{\prime}|$. The direction of movement is consistent with the regulation effect of $t^{\prime}$ on the VHS point shown in Fig.~\ref {Fig1}(d). The situation in Fig.~\ref {Fig9}(b) is very different. With increasing $|t^{\prime}|$, $P_{f_n}$ shows an increasing trend with a potential strong competition with $P_{f}$. The inset of Fig.~\ref {Fig9}(a) also shows the weakening trend of intrinsic pairing interaction $\bm{P}_f$ with larger $|t'|$ and it is similar to the behavior of pairing susceptibility $P_{f}$. Comparing Figs.~\ref {Fig9}(a) and~\ref {Fig9}(b), it is clear that when $|t^{\prime}|$ increases to 0.5, $P_{f_n}$ is generally larger than $P_{f}$ in a wide filling region as $\avg{n}<0.6$. However, as the electronic filling $\avg{n}>0.6$, the inset of Fig.~\ref {Fig9}(a) shows that intrinsic pairing interaction $\bm{P}_{f}$ tends to be lower than zero, and in the inset of Fig.~\ref {Fig9}(b), the value of $\bm{P}_{f_n}$ is always negative for the whole electronic filling range that investigated, even $\bm{P}_{f_n}$ tend to increase with larger $|t'|$.

\section{\romannumeral4. SUMMARY}
For the triangular lattice, the DQMC method shows that the introduction of the next-nearest-neighbor hopping integral $t^{\prime}$ makes the frustration stronger and induces a new competing energy scale, which makes the sign problem more severe. By regulating the shape of the DOS through $t^{\prime}$ and thus the position of the VHS point, we find that a higher DOS favors the formation of ferromagnetic fluctuation. At the same time, the effect of temperature on ferromagnetic fluctuation is significant, and comparative calculations for different lattice sizes show the short-range nature of ferromagnetic correlations and the strengthening effect of the Coulomb interaction $U$ on the ferromagnetic fluctuation at the investigated temperature. To verify this ``strengthening'' effect of $U$, we also compare the energies of different spin subspaces via the CPMC method.

Regarding the superconducting pairing symmetry, the $f$-wave pairing dominates in a rather large filling region. The ground state long-range pairing correlation function of the system is consistent with the behavior of finite temperature pairing susceptibility near the VHS point. In addition, we find that the pairing susceptibility is closely related to the magnetic correlation of the system. Ferromagnetic correlation promotes $f$-wave pairing, and antiferromagnetic correlation tends to favor $f_n$-wave pairing, while the result of simulation at finite temperature shows that both intrinsic pairing interaction $\bm{P_{f}}$ and $\bm{P_{f_n}}$ are negative as $\avg{n}>0.6$. Furthermore, by changing the value of $|t'|$ we regulate the shape of the DOS and thus the position of the VHS point and finally, change the behavior of pairing susceptibility and intrinsic pairing interaction.

\section{Acknowledgements}
This work was supported by Beijing Natural Science
Foundation (No. 1242022) and Guangxi Key Laboratory of Precision Navigation Technology and Application,
Guilin University of Electronic Technology (No. DH202322).

\appendix
\setcounter{equation}{0}
\setcounter{figure}{0}
\renewcommand{\theequation}{A\arabic{equation}}
\renewcommand{\thefigure}{A\arabic{figure}}
\renewcommand{\thesubsection}{A\arabic{subsection}}

\section{Form Factors}
\label{app:falpha}
\begin{figure}[h!]
\includegraphics[width=0.4\textwidth,height=0.3\textwidth]{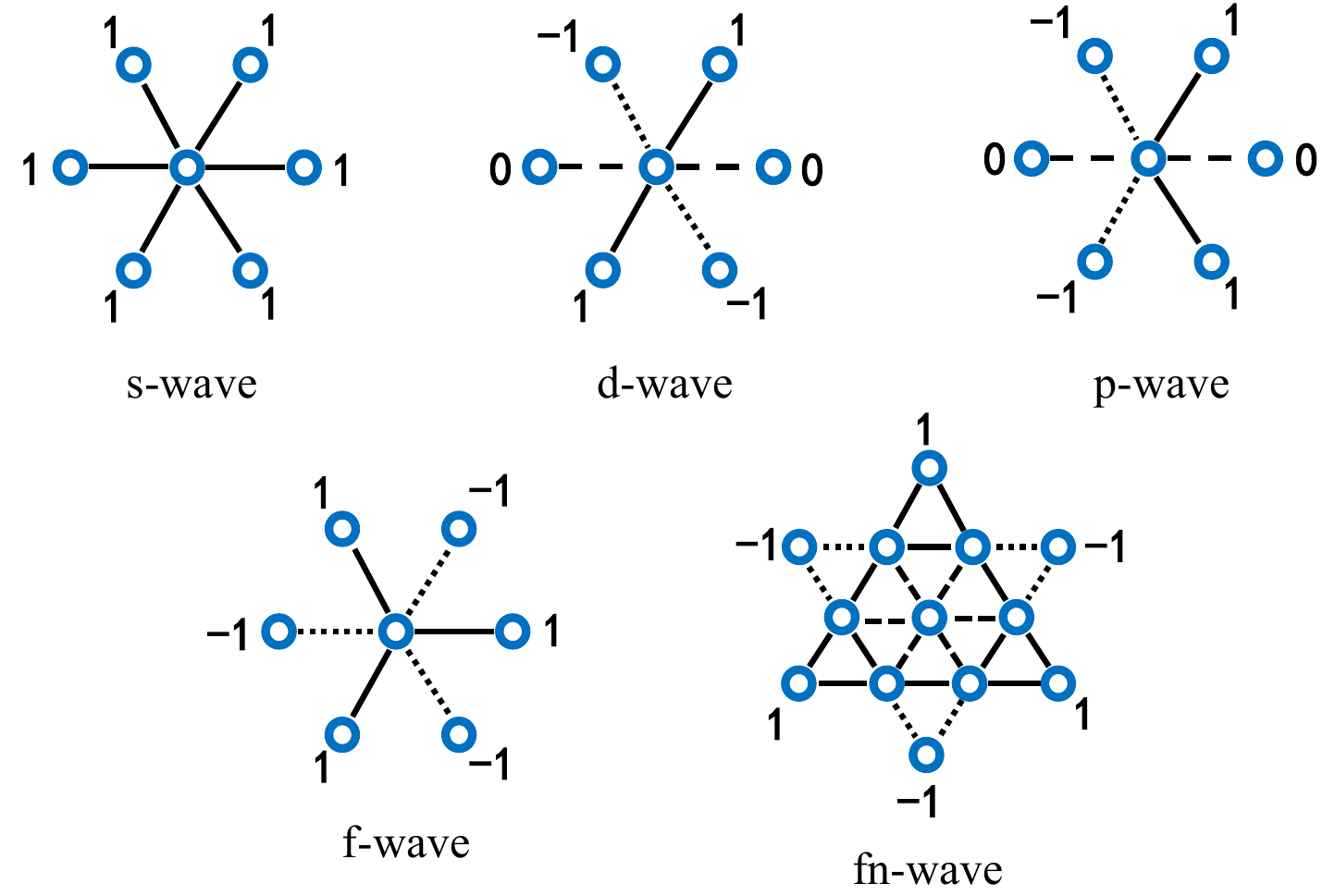}
\caption{\label{Fig5} Site-dependent form factors for the $s$ wave, $d$ wave, $p$ wave, $f$ wave, and $f_n$ wave in the triangular lattice.}
\end{figure}
Considering the symmetry of the triangular lattice, the possible form factors include the following six types\cite{PhysRevB.77.125114}. The first three represent spin-singlet pairings, and the last three represent spin-triplet pairings. Assuming that the triangular lattice we study is isotropic and has no symmetry breaking, the $d_{x^{2}-y^{2}}$ wave and $d_{xy}$ wave as well as $p_{x}$ wave and $p_{y}$ wave are degenerate, and we uniformly express them as the $d$ wave and $p$ wave. The extended $s$ wave and $f$ wave of the next-nearest-neighbor sites are expressed as the $s$ wave and $f_n$ wave, respectively, as shown in Fig.~\ref {Fig5}.
\begin{equation}
\begin{aligned}
f_{s}(l)=&\frac{1}{\sqrt{6}}[\delta_{l,\vec{a}_{1}}+\delta_{l,-\vec{a}_{1}}+\delta_{l,\vec{a}_{2}}+\delta_{l,-\vec{a}_{2}}\\&+\delta_{l,\vec{a}_{1}+\vec{a}_{2}}+\delta_{l,-\vec{a}_{1}-\vec{a}_{2}}],\\
f_{d_{xy}}(l)=&\frac{1}{2}[\delta_{l,\vec{a}_{2}}+\delta_{l,-\vec{a}_{2}}-\delta_{l,\vec{a}_{1}+\vec{a}_{2}}-\delta_{l,-\vec{a}_{1}-\vec{a}_{2}}],\\
f_{d_{x^{2}-y^{2}}}(l)=&\frac{1}{2\sqrt{3}}[2\delta_{l,\vec{a}_{1}}+2\delta_{l,-\vec{a}_{1}}-\delta_{l,\vec{a}_{2}}-\delta_{l,-\vec{a}_{2}}\\&-\delta_{l,\vec{a}_{1}+\vec{a}_{2}}-\delta_{l,-\vec{a}_{1}-\vec{a}_{2}}],\\
f_{f}(l)=&\frac{1}{\sqrt{6}}[\delta_{l,\vec{a}_{1}}-\delta_{l,-\vec{a}_{1}}+\delta_{l,\vec{a}_{2}}-\delta_{l,-\vec{a}_{2}}\\&-\delta_{l,\vec{a}_{1}+\vec{a}_{2}}+\delta_{l,-\vec{a}_{1}-\vec{a}_{2}}],\\
f_{f_n}(l)=&\frac{1}{\sqrt{6}}[\delta_{l,2\vec{a}_{1}+\vec{a}_{2}}-\delta_{l,-2\vec{a}_{1}-\vec{a}_{2}}+\delta_{l,-\vec{a}_{1}+\vec{a}_{2}}\\&-\delta_{l,\vec{a}_{1}-\vec{a}_{2}}-\delta_{l,\vec{a}_{1}+2\vec{a}_{2}}+\delta_{l,-\vec{a}_{1}-2\vec{a}_{2}}],\\
f_{p_{x}}(l)=&\frac{1}{2}[-\delta_{l,\vec{a}_{2}}+\delta_{l,-\vec{a}_{2}}+\delta_{l,\vec{a}_{1}+\vec{a}_{2}}-\delta_{l,-\vec{a}_{1}-\vec{a}_{2}}],\\
f_{p_{y}}(l)=&\frac{1}{2\sqrt{3}}[-2\delta_{l,\vec{a}_{1}}+2\delta_{l,-\vec{a}_{1}}+\delta_{l,\vec{a}_{2}}-\delta_{l,-\vec{a}_{2}}\\&-\delta_{l,\vec{a}_{1}+\vec{a}_{2}}+\delta_{l,-\vec{a}_{1}-\vec{a}_{2}}].
\end{aligned}
\end{equation}

\bibliography{References}
\end{document}